\documentclass[]{iopart}
\usepackage{array}
\usepackage{bm}

\begin{document}

\title{Further developments for the auxiliary field method}

\author{Claude Semay$^1$, Fabien Buisseret$^{1,2}$, Bernard Silvestre-Brac$^3$}
\address{$^1$ Service de Physique Nucl\'{e}aire et Subnucl\'{e}aire, Universit\'{e}
de Mons - UMONS, 20 Place du Parc, 7000 Mons, Belgium}
\address{$^2$ Haute \'Ecole Louvain en Hainaut (HELHa), Chauss\'ee de Binche 159, 7000 Mons, Belgium}
\address{$^3$ LPSC Universit\'{e} Joseph Fourier, Grenoble 1,
CNRS/IN2P3, Institut Polytechnique de Grenoble, 
Avenue des Martyrs 53, 38026 Grenoble-Cedex, France}
\ead{claude.semay@umons.ac.be, fabien.buisseret@umons.ac.be, silvestre@lpsc.in2p3.fr}%


\begin{abstract}
The auxiliary field method is a technique to obtain approximate closed formulae for the solutions of both nonrelativistic and semirelativistic eigenequations in quantum mechanics. For a many-body Hamiltonian describing identical particles, it is shown that the approximate eigenvalues can be written as the sum of the kinetic operator evaluated at a mean momentum $p_0$ and of the potential energy computed at a mean distance $r_0$. The quantities $p_0$ and $r_0$ are linked by a simple relation depending on the quantum numbers of the state considered and are determined by an equation which is linked to the generalized virial theorem. The (anti)variational character of the method is discussed, as well as its connection with the perturbation theory. For a nonrelativistic kinematics, general results are obtained for the structure of critical coupling constants for potentials with a finite number of bound states. 
\end{abstract}

\maketitle

\section{Introduction}
\label{sec:intro}

The auxiliary field method (AFM) is a very powerful method to obtain approximate analytical expressions for the eigenvalues of one, two and many-body systems with both nonrelativistic or semirelativistic kinematics. It has been shown in a series of paper \cite{afm1,afm2,afm3,afmsqrt,afmrelat,afmnbody,afmwf,afmdual} that it can be applied with great success in many physical situations. The basic idea is to replace a problem which is not solvable, for example because of a complicated potential or a semirelativistic kinematics, by another one which can be treated analytically. In so doing, it is necessary to introduce auxiliary fields $\hat{\nu}_k$. The original Hamiltonian $H$ is replaced by a new Hamiltonian $\tilde{H}(\hat{\nu}_k)$, called the AFM Hamiltonian. If these auxiliary fields are chosen as $\hat{\nu}_k(0)$ in order to extremize the AFM Hamiltonian, this one coincides with the original Hamiltonian: $\tilde{H}(\hat{\nu}_k(0))=H$. Thus, both formulations are completely equivalent. The approximation lies in the fact that the auxiliary fields are considered no longer as operators, but as a real constants $\nu_k$. An approximate value of the exact eigenenergy $E$ is then given by an extremal eigenenergy $E(\nu_k(0))$ of the AFM Hamiltonian $\tilde{H}(\nu_k)$, which is in principle much simpler than $H$. An approximate state for the corresponding eigenvalue can also be obtained. The quality of this approximation has been studied and discussed in detail in the papers mentioned above. Among the interesting properties of the AFM, we can mention its great simplicity and its ability to treat on equal footing the ground state and the various excited states. This procedure was first introduced to get rid off the square root kinetic operator in calculations for semirelativistic eigenvalue equations \cite{eiben1,eiben2}. As the AFM is an extension of these first calculations, we just keep the same name for the method.

As it is shown in \cite{afmenv}, the AFM has strong connections with the envelope theory \cite{ET1,ET2,ET3,ET4}. Nevertheless both methods have been introduced from completely different starting points. In particular, the AFM introduces the notion of auxiliary fields which is a key ingredient to interpret the method as a mean field approximation and which can be very useful to compute mean values of observables \cite{afm1,afmnbody}. 

In this work, we present some new and general properties of the AFM. So, only the basic ingredients necessary for the understanding of the subject treated here are recalled in Sect.~\ref{sec:AFM}. We refer the reader to our works mentioned above or to our review paper \cite{afmrev} for an exhaustive overview of the method and its applications. New results or generalizations are presented in the following sections. The connection of the AFM with the generalized virial theorem is presented in Sect.~\ref{sec:virial}. The use of the perturbation theory for the AFM is explained in Sect.~\ref{sec:pert}. For a nonrelativistic kinematics, the general structures of critical coupling constants for potentials with a finite number of bound states are presented in Sect.~\ref{sec:critic}. A summary of the results is given in Sect.~\ref{sec:sum}.

\section{The auxiliary field method}
\label{sec:AFM}

Let us consider a system composed of $N$ particles, interacting via one-body potentials $U_i$ and two-body potentials $V_{ij}$, and moving with a nonrelativistic or a semirelativistic kinetic energy. In principle, the  AFM can treat this problem in such a general form, but it is manageable in practice only if the particles are identical. This implies that they all have the same mass $m$, that the form of the one-body potentials is the same for all particles $U_i \equiv U$, and that the form of the two-body potentials is the same for all pairs of particles $V_{ij} \equiv V$. So, the most general Hamiltonian we will consider in this paper has the following form
\begin{equation}
\label{HN}
H = \sum_{i=1}^N \sqrt{\bm p_i^2 + m^2} + \sum_{i=1}^N U(|\bm s_i|) +
\sum_{i<j = 1}^N V(|\bm r_{ij}|),
\end{equation}
with $\bm s_i=\bm r_i - \bm R$ and $\bm r_{i j}=\bm r_i - \bm r_j$. $\bm r_i$ is the position of the particle $i$, $\bm p_i$ is its conjugate momentum and $\bm R$ is the position of the center of mass of the $N$ particles ($N \ge 2$). It is assumed that $\sum_{i=1}^N \bm p_i = \bm 0$. Following the AFM, this Hamiltonian is ``replaced" by the auxiliary Hamiltonian $\tilde H$, with auxiliary potentials $P(x)$ and $S(x)$, and which depends on 3 auxiliary fields $\mu$, $\nu$ and $\rho$
\begin{equation}
\label{HNtilde}
\tilde H(\mu,\nu,\rho)= B(\mu,\nu,\rho)+\sum^{N}_{i=1}\frac{\bm p^2_i}{2\mu}+\nu\sum^{N}_{i=1}P(|\bm s_i|)+
\rho\sum^{N}_{i<j=1} S(|\bm r_{ij}|), 
\end{equation}
provided the states considered are completely symmetrized. Most of the results presented in this section come from \cite{afmnbody}, but we use here the more convenient notations developed in \cite{afmdual,afmrev}. The function $B(\mu,\nu,\rho)$ is not useful to detail in this work, but it can be rebuilt from results in \cite{afmnbody}. If this Hamiltonian is analytically solvable, an AFM analytical approximation of a mass $M$ of the $N$-body Hamiltonian $H$ is given by an eigenvalue $M_0$ of the Hamiltonian $\tilde H(\mu_0,\nu_0,\rho_0)$ where the 3 optimal auxiliary parameters $\mu_0$, $\nu_0$ and $\rho_0$ extremize this eigenvalue. These parameters depend on the quantum numbers of the state.

At this stage, it is not obvious that the solution $M_0$ is a good one. But, the comparison theorem of the quantum mechanics can be used to obtain significant information about the AFM eigenvalues. This theorem states that, for some eigenvalue equations, if two Hamiltonians are ordered, $H^{(1)} \le H^{(2)}$ ($\langle H^{(1)} \rangle \le \langle H^{(2)} \rangle$ for any state), then each corresponding pair of eigenvalues is ordered $E^{(1)}_{\{\theta\}} \le E^{(2)}_{\{\theta\}}$, where ${\{\theta\}}$ represents a set of quantum numbers. This inequality can be obtained from the Ritz variational principle \cite{reedsimon}, but it can also be derived from the Hellmann-Feynman theorem \cite{comptheor}. If we can show that the auxiliary Hamiltonian $\tilde H(\mu_0,\nu_0,\rho_0)$ is greater or lower than the genuine Hamiltonian $H$, then it is possible to use the comparison theorem to locate the AFM eigenvalues with respect to the exact ones. 

In the case of a nonrelativistic kinematics, The AFM yield an upper (lower) bound if the potentials $U(x)$ and $V(x)$ could be bounded from above (below) by the auxiliary potentials $P(x)$ and $S(x)$ respectively \cite{afmnbody}. For a semirelativistic kinematics, the AFM implies a replacement of the square root operators by a nonrelativistic form of the kinetic energy (see (\ref{HNtilde})) and this yields to an increase of the eigenvalues \cite{afmrelat,comptheor}. So, in this case, the AFM solutions are upper bounds of the exact ones if the potentials $U(x)$ and $V(x)$ can be bounded from above by the auxiliary potentials $P(x)$ and $S(x)$ respectively. In other cases, nothing can be said about the possible variational character of the solutions. 

Let us note that a lower bound for the ground state (and then for the whole spectrum) of the general Hamiltonian (\ref{HN}) for a boson-like system \footnote{A boson-like system is composed of particles whose total spatial wavefunction can be completely symmetrical. For instance, this is the case for a system of quarks inside a baryon: quarks are fermions, but the baryon is characterized by a completely antisymmetrical colour function so that the rest of the total wavefunction must be completely symmetrical. Similarly, a fermion-like system is composed of particles whose total spatial wavefunction can be completely antisymmetrical.} has been proposed in \cite{lowerb1}. It takes the following general form (the one-body potential is introduced using the fact that $|\bm s_1| = |\bm s_2| = |\bm r_{12}|/2$ for $N=2$)
\begin{equation}
\label{lowerb}
M \ge N \inf_{\phi} \left\langle \phi \left| \sqrt{\bm p^2 + m^2} +  U\left(\frac{1}{2}|\bm r|\right)+ \frac{N-1}{2} V(|\bm r|) \right| \phi \right\rangle,
\end{equation}
but also works for nonrelativistic kinematics \cite{lowerb2}. In this latter case, using the AFM results with $N=2$, a lower bound for the mean value in (\ref{lowerb}) can be computed provided the potentials $U(x)$ and $V(x)$ could be bounded from below by the auxiliary potentials (see Sect.~\ref{sec:AFMN12}). 

\subsection{The case $N\ge 2$}
\label{sec:AFMNgt2}

For arbitrary values of $N$, the Hamiltonian (\ref{HNtilde}) is entirely analytically solvable for the unique choice $P(x)$ and/or $S(x)$ equal to $x^2$. It can then be shown \cite{afmnbody} that the non linear system determining the 3 variables $(\mu_0,\nu_0,\rho_0)$ can be recast in the form of one transcendental equation depending on the single variable $X_0 = \sqrt{2 \mu_0 (\nu_0 + N \rho_0)}$. Moreover, an eigenmass can be computed from the $X_0$ quantity only. Thus, the eigenvalue problem for the $N$-body system can be determined simply by the set of the following two equations: \cite{afmnbody} 
\begin{eqnarray}
\label{MX0}
&& \fl M_0=N \sqrt{m^2+\frac{Q}{N}X_0} + N U \left( \sqrt{\frac{Q}{NX_0}} \right) + C_N
       V \left( \sqrt{\frac{2Q}{(N-1)X_0}} \right), \\
\label{X0}
&& \fl X_0^2 = 2 \sqrt{m^2+\frac{Q}{N}X_0} \left[K \left( \sqrt{\frac{Q}{NX_0}} \right)
        + N L \left( \sqrt{\frac{2Q}{(N-1)X_0}} \right) \right],
\end{eqnarray}
where $Q$ is a global quantum number (see below), where $K(x)=U'(x)/P'(x)=U'(x)/(2 x)$ and $L(x)=V'(x)/S'(x)=V'(x)/(2 x)$, and where the number of pairs
\begin{equation}
\label{CN}
C_N = \frac{N(N-1)}{2}
\end{equation}
has been introduced for convenience. The prime denotes the derivative with respect to the argument. In this framework, an approximate AFM eigenstate is given by an eigenstate of $\tilde H(\mu_0,\nu_0,\rho_0)$. It is written in terms of Jacobi coordinates as a product of $(N-1)$ oscillator states with sizes depending on $N$ and $X_0$ \cite{afmnbody}. A nonrelativistic version of (\ref{MX0})-(\ref{X0}) can be obtained in the limit $m\to\infty$ \cite{afmnbody}. In this case, $\mu_0 \to m$. Further simplifications occur also for the ultrarelativistic limit $m=0$. 

A state depends on $(N-1)$ radial quantum numbers $n_i$ and $(N-1)$ orbital quantum numbers $l_i$, as well as intermediate coupling quantum numbers which are not considered here. The global quantum number resulting from the AFM treatment is then 
\begin{equation}
\label{Q}
Q = \sum_{i=1}^{N-1} (2 n_i + l_i) + \frac{3}{2} (N - 1).
\end{equation}
All quantum numbers are not allowed, depending on the nature of the particles. In particular, the ground state for a boson-like system is just $Q = 3(N-1)/2$, while the ground state of a fermion-like system is much more involved and needs the introduction of the Fermi level \cite{afmnbody}. 

Since $P(x)$ and/or $S(x)$ equal to $x^2$ can only be used, an upper bound is computed for most of the relevant interactions, a fortiori for a semirelativistic kinematics. For instance, an AFM mass formula has been obtained for a system of $N$ relativistic massless quarks interacting via a linear one-body confinement and a two-body Coulomb potential (this kind of Hamiltonian is pertinent for variant theories of the quantum chromodynamics). The accuracy of this formula has been numerically tested in \cite{afmnbody} with $N=3$: Relative errors less than 20\% have been obtained for the lowest states. It has also be shown in \cite{baryonlnc} that the $N$-dependence of this formula is the correct one for $N\to\infty$. When a closed formula cannot be computed, numerical solutions (generally upper bounds) can always be easily obtained. This is valuable for a $N$-body system. 

\subsection{The cases $N=1$ and $N=2$}
\label{sec:AFMN12}

For $N=2$, $\bm s_1 = -\bm s_2 = \bm r_{12}/2$. So, the potential $U(x)$ becomes redundant with the potential $V(x)$ and can be ignored. Moreover, the Hamiltonian $H$ simplifies because $\bm p_1 = -\bm p_2= \bm p$. Thus, for a nonrelativistic kinematics, the case of two different particles can be considered by replacing the kinetic part $2 m+\bm p^2/m$ by $m_1+m_2+\bm p^2/(2 m_r)$ where $m_r$ is the reduced mass. A priori, above calculations are only valid for $N\ge 2$. But, starting from the one-body equivalent of Hamiltonian (\ref{HN}), it can be shown that equations (\ref{MX0})-(\ref{X0}) are also relevant for $N=1$ by setting $V(x)=0$ and reinterpreting $\bm p_1$ and $\bm s_1$ as conjugate variables. 

For both $N=1$ and 2 systems, the more general form $\textrm{sgn}(\lambda)\,x^\lambda$ can be used for the auxiliary potential, instead of only $x^2$. This leads to various expressions for $Q$. The complete calculation shows that the same system (\ref{MX0})-(\ref{X0}) is found and that the only trace of the auxiliary potential lies in the structure of the global quantum number $Q$. In practice, $Q=2 n+l+3/2$ with $P(x)$ or $S(x)=x^2$ (see (\ref{Q}) with $N=2$), $Q=n+l+1$ with $P(x)$ or $S(x)=-1/x$ \cite{afm3} and $Q=2 ( -\alpha_n/3 )^{3/2}$ for S-wave states with $P(x)$ or $S(x)=x$ \cite{afmwf}, where $\alpha_n$ is the $(n+1)^{\textrm{th}}$ zero of the Airy function Ai. Depending on the kinematics, closed form formulae have been obtained for various potentials: sum of two power-law, logarithmic, Yukawa, exponential, square-root \cite{afm1,afm2,afm3,afmsqrt,afmrelat,afmnbody}. If a closed formula cannot be computed, the method is then not really interesting since a lot of numerical techniques can be harnessed to find accurate solutions for one- or two-body systems.
 
For some nonrelativistic systems, it is possible to use two forms of the auxiliary potential to obtain both upper and lower analytical bounds of the exact solutions. The following potentials, $a \sqrt{x^2+b^2}$, $a \ln (b x)$, $a\, x-b/x$, or $\textrm{sgn}(\lambda)\,a x^\lambda$ (with $a>0$, $b>0$, $-1 \le \lambda \le 2$) can be bounded from below (above) with the choice $-1/x$ ($x^2$) for the auxiliary potential. For instance, let us consider the Hamiltonian
\begin{equation}
H=\frac{\bm p^2}{2 \mu}+\sqrt{a^2 r^2+b^2}.
\end{equation}
The eigenenergies computed with the AFM gives
\begin{equation}
\fl
E_\textrm{AFM}=\frac{2 b}{\sqrt{3 Y}} \left(G_-^2(Y) + \frac{1}{G_-(Y)}\right) \quad \textrm{with} \quad 
Y=\frac{b^2}{3}\left(\frac{32 \,\mu}{a^2Q^2}\right)^{2/3},
\end{equation}
and where $G_-(Y)$ is the solution of the equation $4\,G_-(Y)^4-8\,G_-(Y)-3\,Y=0$. Upper (lower) bounds are obtained with $Q=2n+l+3/2$ ($Q=n+l+1$). The quality of these bounds are studied in \cite{afmsqrt,afmrelat} where more details are given about this solution and the function $G_-(Y)$. 

\section{Connection with the virial theorem}
\label{sec:virial}

The general virial theorem links the mean values of the directional derivatives of the kinetics operator and the potential \cite{virial1,virial2}. Using the Hellmann-Feynman theorem as in \cite{virial3}, it can be applied to the general $N$-body Hamiltonian (\ref{HN}) to yield
\begin{equation}
\label{virial}
\fl
N \left\langle {\bm p_k}\cdot\bm\nabla_{\bm p_k} T(\bm p_k) \right\rangle = N \left\langle {\bm s_l}\cdot\bm\nabla_{\bm s_l} U(\bm s_l) \right\rangle + C_N \left\langle {\bm r_{i j}}\cdot\bm\nabla_{\bm r_{i j}} V(\bm r_{i j}) \right\rangle,
\end{equation}
with arbitrary numbers \{$k$, $l$, $i\ne j$\} if the mean values are taken with a completely symmetrized eigenstate of the $N$-body Hamiltonian. The operator $T$ is defined by $T(\bm x)= \sqrt{\bm x^2+m^2}$ or by its nonrelativistic counterpart $m+\bm x^2/(2m)$. Let us introduce the distance $r_0=\sqrt{N Q/X_0}$ and the momentum $p_0=Q/r_0$. It is a simple algebra exercise to show that formulae (\ref{MX0})-(\ref{X0}) can be written as:
\begin{eqnarray}
\label{M0}
&&M_0=N\, T(p_0) + N\, U \left( \frac{r_0}{N} \right) + C_N\, V \left( \frac{r_0}{\sqrt{C_N}} \right), \\
\label{pQr}
&&p_0=\frac{Q}{r_0}, \\
\label{p0r0}
&&N\, p_0 T'(p_0) =  N\, \frac{r_0}{N}U' \left( \frac{r_0}{N} \right) + C_N\, \frac{r_0}{\sqrt{C_N}}V' \left( \frac{r_0}{\sqrt{C_N}} \right).
\end{eqnarray}
These equations have not been presented in our previous papers. Before discussing their physical meaning, let us look at the quantities $r_0$ and $p_0$. Using formulae of the appendixes in \cite{afmwf} and \cite{baryonlnc}, the following observables can be analytically computed:
\begin{eqnarray}
\label{p0}
&&\frac{1}{N}\left\langle \sum_{i=1}^N \bm p_i^2 \right\rangle = p_0^2, \\
\label{r0}
&&N \left\langle \sum_{i=1}^N \bm s_i^2 \right\rangle =
\left\langle \sum_{i<j=1}^N \bm r_{ij}^2 \right\rangle = r_0^2. 
\end{eqnarray}
This shows that $r_0$ can be considered as a mean radius for the system and $p_0$ as a mean momentum per particle. Indeed, (\ref{p0}) and (\ref{r0}) imply that
\begin{equation}
\label{meanvar}
\sqrt{\left\langle\bm p_i^2 \right\rangle}=p_0, \quad
\sqrt{\left\langle\bm s_i^2 \right\rangle}=\frac{r_0}{N}, \quad
\sqrt{\left\langle\bm r_{ij}^2 \right\rangle}=\frac{r_0}{\sqrt{C_N}},
\end{equation}
for arbitrary $i\ne j$ since the mean values are taken with completely symmetrized states. These results can also be obtained using the more general relations (66)-(68) in \cite{afmnbody} relevant for $P(x)$ and $S(x)$ different from $x^2$.

With this new formulation, an AFM eigenvalue given by (\ref{M0}) is simply the kinetic operator evaluated at the mean momentum $p_0$ plus the potential energy computed at some mean radius depending on $r_0$. As one could expect, the kinetic energy and the one-body potential energy are proportional to the number of particles and the two-body potential energy is proportional to the number of pairs. Formula (\ref{M0}) looks like a semiclassical approximation but this is absolutely not the case. The AFM yields an approximate $N$-body wavefunction \cite{afmnbody,baryonlnc}, and the relation (\ref{pQr}) between $p_0$ and $r_0$ is a full quantum link, function of the quantum numbers of the system. At last, the value of $r_0$ (and thus of $p_0$) is the solution of a transcendental equation (\ref{p0r0}) which is the translation into the AFM variables of the generalized virial theorem (\ref{virial}) which comes from very general properties of quantum mechanics. These considerations prove that the AFM really relies on very sound physical basis. Once the system (\ref{M0})-(\ref{p0r0}) is written, it can appear finally quite natural to obtain such a result. The problem is to find a relevant link between the mean values $r_0$ and $p_0$. This is solved by the AFM.

It is generally possible to improve the quality of the AFM eigenvalues with a slight modification of the principal quantum number. A particularly simple form which works quite well is given by
\begin{equation}
\label{Qmod}
Q = \sum_{i=1}^{N-1} (\alpha\, n_i + \beta\, l_i) + \gamma(N-1),
\end{equation}
where the values of parameters $\alpha$, $\beta$ and $\gamma$ depend on both the interaction and the kinematics. They can be determined by an analytical procedure in some cases by using analytical results coming from WKB approximations or variational calculations \cite{afm1,afmdual}. Even if it less interesting, a fit on numerically computed exact eigenvalues can always be implemented \cite{afm1,afm2,afm3,afmsqrt,afmrelat,afmnbody}. With the form (\ref{Qmod}), the variational character of the AFM approximation is lost, but the relative errors can be sometimes strongly reduced. 

\section{Connection with the perturbation theory}
\label{sec:pert}

It has been shown in \cite{afm2} that, for one- and two-body nonrelativistic systems, the AFM and the perturbation theory give similar results when the potential is an exactly solvable one plus a small perturbation. This result is extended here for the general Hamiltonian (\ref{HN}), that is to say: $N$ particles, semirelativistic kinematics and arbitrary potentials $U(x)$ and $V(x)$. 

Let us first assume that each pairwise potential $V(|\bm r_{ij}|)$ is supplemented by a term $\epsilon \, v(|\bm r_{ij}|)$, with $\epsilon \ll 1$ in order that $\epsilon \, v(x) \ll V(x)$ in the physical domain of interest. In the system (\ref{M0})-(\ref{p0r0}), the potential $V(x)$ is replaced by $V(x)+\epsilon \, v(x)$. In this case, new values $r_1$  and $p_1$ for the mean radius and momentum will be the solution of the new system:
\begin{eqnarray}
\label{M1}
&& \fl M_1=N\, T(p_1) + N\, U \left( \frac{r_1}{N} \right) + C_N \left[V \left( \frac{r_1}{\sqrt{C_N}} \right)+ \epsilon\, v \left( \frac{r_1}{\sqrt{C_N}} \right) \right], \\
\label{pQr1}
&& \fl p_1\, r_1=Q, \\
\label{p1r1}
&& \fl N\, p_1 T'(p_1) = r_1 U' \left( \frac{r_1}{N} \right) + \sqrt{C_N}\, r_1 \left[ V' \left( \frac{r_1}{\sqrt{C_N}} \right)+ \epsilon\,v' \left( \frac{r_1}{\sqrt{C_N}} \right) \right].
\end{eqnarray} 
Writing $r_1 = (1+\delta)r_0$, we can expect $\delta \ll 1$ since $\epsilon \ll 1$. In this case, power expansions at first order can be computed. We have $p_1 \approx (1-\delta)p_0$ from (\ref{pQr1}), and we can write $T(p_1)\approx T(p_0)-\delta\, p_0\, T'(p_0)$, $T'(p_1)\approx T'(p_0)-\delta\, p_0\, T''(p_0)$, $U(r_1/N)\approx U(r_0/N)+ \delta\, r_0\, U'(r_0/N)/N$, etc. Equation (\ref{p1r1}) reduces to an expression of the form $\delta \approx \epsilon\, h(r_0)$ where $h$ is a quite complicated function of $T'$, $U'$, $V'$ and their derivatives. It confirms that $\delta \sim \Or (\epsilon)$. The precise form of $h$ is given below in the most general case. It is then possible to perform an expansion of $M_1$ to obtain
\begin{eqnarray}
\label{M1exp}
&& \fl M_1=N\, T(p_0) - N\, \delta\, p_0 T'(p_0)
+ N\, U \left( \frac{r_0}{N} \right) + \delta\, r_0 U' \left( \frac{r_0}{N} \right) \nonumber \\
&& \fl + C_N\, V \left( \frac{r_0}{\sqrt{C_N}} \right)+ \sqrt{C_N}\, \delta\, r_0 V' \left( \frac{r_0}{\sqrt{C_N}} \right) 
+ C_N\, \epsilon\, v \left( \frac{r_0}{\sqrt{C_N}} \right) + \Or (\epsilon^2).
\end{eqnarray}
Using (\ref{M0}) and (\ref{p0r0}), this equations simplifies to
\begin{equation}
\label{M1expv}
M_1=M_0 
+ C_N\, \epsilon\, v \left( \frac{r_0}{\sqrt{C_N}} \right) + \Or (\epsilon^2).
\end{equation}
This result could seem quite obvious, but it demonstrates that the knowledge of $r_0$ is sufficient to obtain the contribution of the perturbation at the first order.

Let us now look at the most general case and assume too that each [one-body potential $U(|\bm s_i|)$ / kinetic operator $T(|\bm p_i|)$] is supplemented by a term [$\eta \, u(|\bm s_i|)$ / $\tau \, t(|\bm p_i|)$], with [$\eta \ll 1$ / $\tau \ll 1$] in order that [$\eta \, u(x) \ll U(x)$ / $\tau \, t(x) \ll T(x)$] in the physical domain of interest. With similar calculations, we finally find
\begin{equation}
\label{M1tot}
\fl
M_1=M_0 
+ N\, \tau\, t \left( p_0 \right)
+ N\, \eta\, u \left( \frac{r_0}{N} \right) 
+ C_N\, \epsilon\, v \left( \frac{r_0}{\sqrt{C_N}} \right)
+ \Or (\epsilon^2,\eta^2,\tau^2).
\end{equation}
The parameter $\delta$ is determined at the same order by the following relation
\begin{eqnarray}
&& \fl N\, p_0\, \tau\, t' \left( p_0 \right)
- r_0\, \eta\, u' \left( \frac{r_0}{N} \right) 
- \sqrt{C_N}\, r_0\, \epsilon\, v' \left( \frac{r_0}{\sqrt{C_N}} \right)  \nonumber\\
&& \fl =\delta 
\left[ 2\, N\, p_0\, T' \left( p_0 \right) 
+ N\, p_0^2\, T'' \left( p_0 \right)
+ \frac{r_0^2}{N} U''\left( \frac{r_0}{N} \right)
+ r_0^2\, V'' \left( \frac{r_0}{\sqrt{C_N}} \right)\right].
\end{eqnarray}
Perturbed observables and wavefunctions can then be computed at first order, since $r_1 = (1+\delta)r_0$ and $p_1 = (1-\delta)p_0$ at this order.

The contribution of a perturbation at the first order can thus be very easily computed within the AFM once the unperturbed problem is solved. In order to check the quality of this approximation, let us consider a case in which the unperturbed Hamiltonian $H$ can be solved exactly by the AFM, that is $M_0$ is the exact solution. If the small perturbation potential is written $\epsilon \sum_{i<j=1}^N v(|\bm r_{ij}|)$, the quantum perturbation theory says that the solution $M_*$ is given by
\begin{equation}
\label{Mpert}
M_*=M_0 + C_N\, \epsilon\, \langle v(|\bm r_{ij}|) \rangle + \Or (\epsilon^2),
\end{equation}
for any pair $(ij)$. The mean value is taken with a completely symmetrized eigenstate of the unperturbed Hamiltonian $H$. The comparison of (\ref{Mpert}) with (\ref{M1expv}) shows that $\langle v(|\bm r_{ij}|) \rangle$ is replaced by $v \left( r_0/\sqrt{C_N} \right)$ within the AFM. This is to be compared with the exact relation $\langle S(|\bm r_{ij}|) \rangle = S \left( r_0/\sqrt{C_N} \right)$ for the auxiliary potential \cite{afm1,afmnbody}. So, the AFM does not give the same result as the perturbation theory. But the agreement can be very good, as shown with several examples calculated explicitly in \cite{afm2}. Similar discussions can be made for small one-body perturbation potentials or small perturbations of the kinematics. 

\section{Critical coupling constants}
\label{sec:critic}

Some interactions, as the Yukawa or the exponential potentials, admit only a finite number of bound states. Let us assume that such an interaction can be written as $W(x)=-\kappa\, w(x)$, where $\kappa$ is a positive quantity which has the dimension of an energy and $w(x)$ a ``globally positive" dimensionless function such that $\lim_{x\to\infty} w(x) = 0$. We can introduce the notion of critical coupling constant $\kappa(\{\theta\})$ where $\{\theta\}$ stands for a set of quantum numbers. This quantity is such that, if $\kappa > \kappa(\{\theta\})$, the potential admits a bound state with the quantum numbers $\{\theta\}$. The interaction energy for the state with quantum numbers $\{\theta\}$ is then just vanishing for $\kappa = \kappa(\{\theta\})$. We refer the reader to \cite{brau1,brau2,brau3,brau4} for detailed explanations about how to compute critical coupling constant in a given potential.

Let us consider a nonrelativistic $N$-body system (no manageable calculation can be performed for a semirelativistic kinematics) with one-body potentials $U(x)=-k\, u(x)$ and two-body potentials $V(x)=-g\, v(x)$, both independent of the particle mass and both admitting only a finite number of bound states. The system (\ref{M0})-(\ref{p0r0}) for a vanishing energy gives:
\begin{eqnarray}
\label{E0a}
N\, \frac{Q^2}{2\ m\, r_0^2} &= &N\,k_N\, u \left( \frac{r_0}{N} \right) + C_N\,g_N\, v \left( \frac{r_0}{\sqrt{C_N}} \right), \\
\label{E0b}
N\, \frac{Q^2}{m\, r_0^2} &=&-k_N\, r_0\, u' \left( \frac{r_0}{N} \right) - \sqrt{C_N}\,g_N\, r_0\, v' \left( \frac{r_0}{\sqrt{C_N}} \right),
\end{eqnarray}
where $k_N$ and $g_N$ are the critical constants for the system with $N$ particles. The elimination of the ratio $N\, Q^2/(m\, r_0^2)$ from both equations yields the equality
\begin{eqnarray}
\label{E0ab}
&&2 N\,k_N\, u \left( \frac{r_0}{N} \right) + 2 C_N\,g_N\, v \left( \frac{r_0}{\sqrt{C_N}} \right)\nonumber \\
&&=-k_N\, r_0\, u' \left( \frac{r_0}{N} \right) - \sqrt{C_N}\,g_N\, r_0\, v' \left( \frac{r_0}{\sqrt{C_N}} \right).
\end{eqnarray}
When potentials $u$ and $v$ are both taken into account, nothing interesting can be said. So let us consider one type of potential at once. 

Assuming that only two-body forces are present, (\ref{E0ab}) reduces to
\begin{equation}
\label{E0abg}
2\, \sqrt{C_N}\, v \left( \frac{r_0}{\sqrt{C_N}} \right) + r_0\, v' \left( \frac{r_0}{\sqrt{C_N}} \right) =0,
\end{equation}
where the parameter $g_N$ has disappeared. Introducing the new variable $y_0=r_0/\sqrt{C_N}$, we can rewrite 
(\ref{E0a}) and (\ref{E0b}) as:
\begin{eqnarray}
\label{gQ}
&&g_N  = \frac{1}{y_0^2\,v(y_0)} \frac{2}{N(N-1)^2} \frac{Q^2}{m}, \\
\label{y0gQ}
&&2\, v(y_0)+y_0\, v'(y_0)=0.
\end{eqnarray}
The variable $y_0$, determined by (\ref{y0gQ}), is independent of $N$, $Q$ and $m$, and depends only on the form of the function $v(x)$. So, the general formula (\ref{gQ}), which was not obtained in our previous works, gives precise information about the dependence of the many-body critical coupling constant $g_N$ as a function of all the characteristics of the system. With the system (\ref{gQ})-(\ref{y0gQ}), it is easy to recover some limited previous AFM results obtained for the critical coupling constants of Yukawa and exponential interactions \cite{afm3,afmnbody}. For instance, with the two-body Yukawa interaction $V(x)=-g \exp(-\beta x)/x$, we have 
\begin{equation}
\label{gyuk}
g_N  = \frac{2\, e\, \beta\, Q^2}{N(N-1)^2 m}.
\end{equation}
For $N=2$ and $Q=n+l+1$, reasonable upper bounds of the exact critical coupling constants are obtained \cite{afm3}.

Within the AFM approximation, the ground state (GS) of a boson-like system is characterized by $Q=\frac{3}{2}(N-1)$. We obtain in this case the following very general relation valid, at the AFM approximation, for all pairwise potentials with a finite number of bound states
\begin{equation}
\label{gnnp1}
\frac{g_{N+1}(\textrm{GS})}{g_N(\textrm{GS})}=\frac{N}{N+1}.
\end{equation}
This ratio has previously been obtained and numerically checked for several exponential-type potentials \cite{rich94,mosz00}. Similarly, in the same general situation, 
\begin{equation}
\label{gn2}
g_N(\textrm{GS})=\frac{2}{N}g_2(\textrm{GS}),
\end{equation}
indicating that in order to bind a $N$-body system, a coupling $N/2$ times smaller than the coupling for a two-body problem is sufficient \cite{rich94,mosz00}.

Assuming that only one-body forces are present, a similar calculation gives:
\begin{eqnarray}
\label{kQ}
&& k_N = \frac{1}{y_0^2\,u(y_0)} \frac{1}{2 N^2} \frac{Q^2}{m},
\\
\label{y0kQ}
&&2\, u(y_0)+y_0\, u'(y_0)=0,
\end{eqnarray}
where the change of variable $y_0=r_0/N$ has been used. Again, the general formula (\ref{kQ}), which was not obtained in our previous works, gives precise information about the dependence of the one-body critical coupling constant $k_N$ as a function of all the characteristics of the system. For the ground state of a boson-like system, we obtain: 
\begin{eqnarray}
\label{knnp1}
\frac{k_{N+1}(\textrm{GS})}{k_N(\textrm{GS})}&=&\left(\frac{N^2}{N^2-1}\right)^2, \\
\label{kn2}
k_N(\textrm{GS})&=&4\left(\frac{N-1}{N}\right)^2 k_2(\textrm{GS}).
\end{eqnarray}
These results are strongly different from those for pairwise forces. 

If the AFM gives upper (lower) bounds for the exact eigenvalues, the critical coupling constants predicted by formulae above are upper (lower) bounds for the exact critical coupling constants. 

\section{Summary}
\label{sec:sum}

The main interest of the auxiliary field method is to obtain approximate closed formulae for the solutions of  nonrelativistic and semirelativistic eigenequations in quantum mechanics. The idea, strongly connected with the envelope theory, is to replace a Hamiltonian $H$ for which analytical solutions are not known by another one $\tilde H$ which is solvable and which includes one or more auxiliary real parameters. The approximant solutions for $H$, eigenvalues and eigenfunctions, are then obtained by the solutions of $\tilde H$ in which the auxiliary parameters are eliminated by an extremization procedure for the eigenenergies. The AFM can yield upper or lower bounds (both in some favorable situations) on the exact eigenvalues. The nature of the bound depends on the fact that $\tilde H  \ge H$ or $\tilde H \le H$. With a semirelativistic kinematics, only upper bounds can be obtained because of the replacement of the kinetic operator by a nonrelativistic one. For many-body systems, only one type of Hamiltonian $\tilde H$ can be used. So, it is not possible to obtain both upper and lower bounds for the whole spectrum in this case. Nevertheless, for a nonrelativistic kinematics, a lower bound for the ground state can be sometimes computed. 

Provided the structure of the Hamiltonian $\tilde H$ is well chosen (nonrelativistic kinematics plus power-law potentials), an eigenvalue computed by the AFM is simply the kinetic operator evaluated at a mean momentum $p_0$ plus the potential energy computed at some functions of the mean radius $r_0$. The product $r_0\,p_0$ is equal to a global quantum number characterizing the state considered, and the value of $r_0$ (and then of $p_0$) is the solution of a transcendental equation which is the translation into the AFM variables of the generalized virial theorem. This new result gives sound physical basis to the method. 

Once a problem is solved within the AFM, it is very easy to compute the contribution of a small perturbation at the first order. It is given by the perturbation Hamiltonian evaluated at the mean momentum $p_0$ for a kinetic energy or at a function of the mean radius $r_0$ for a potential. The result does not coincide with the one obtained by the quantum perturbation theory, but the agreement can be very good. 

The AFM gives a very general formula for the critical coupling constants of nonrelativistic Hamiltonians with a finite number of bound states. The dependence on the quantum numbers, the mass $m$ of the particles, the number $N$ of particles, and the structure of the potential are predicted. Different $N$ behaviours are obtained depending on the one-body or pairwise character of the interaction. If the AFM gives upper (lower) bounds for the exact eigenvalues, the critical coupling constants predicted are upper (lower) bounds for the exact critical coupling constants. 

\ack
C. Semay and F. Buisseret would thank the F.R.S.-FNRS for financial support.

\end{document}